\documentstyle[aps,prl,amssymb,latexsym,epsfig]{revtex}

\newcommand{\beq}{\begin{equation}}
\newcommand{\eeq}{\end{equation}}
\newcommand{\bqa}{\begin{eqnarray}}
\newcommand{\eqa}{\end{eqnarray}}
\newcommand{\fr}{\frac}

\begin{document}

\draft

\wideabs{
\title{Shell crossing in generalized Tolman-Bondi spacetimes}
\author{S\' ergio M. C. V. Gon\c calves}
\address{Theoretical Astrophysics, California Institute of Technology, Pasadena, California 91125}
\date{August 28, 2000}
\maketitle
\begin{abstract}
We study the occurrence of shell crossing in spherical weakly charged dust collapse in the presence of a non-vanishing cosmological constant. We find that shell crossing always occurs from generic time-symmetric regular initial data, near the center of the matter configuration. For non-time-symmetric initial data, the occurrence---or lack thereof---of shell crossing is determined by the initial velocity profile, for a given mass and charge distribution.  Physically reasonable initial data inevitably leads to shell crossing (near the center) before the minimum bounce radius is reached.
\end{abstract}
}

\narrowtext

\section{INTRODUCTION}

Tolman-Bondi metrics \cite{tolman34,bondi48} describe the gravitational collapse of spatially bounded spherical dust configurations in an otherwise empty spacetime. The model consists of a general spherically symmetric metric, matched to a Schwarzschild exterior, with the matter content being general inhomogeneous dust. Spherical symmetry implies that the initial data---two arbitrary functions giving the initial density and velocity profiles---are a function of the radial coordinate $r$ alone. Since the collapse is pressureless, every shell of dust with finite proper radius will collapse through its Schwarzschild radius, become trapped and proceed to become singular at the center of symmetry in a finite amount of proper time (as measured by an observer comoving with the shell).

For homogeneous dust distributions---Oppenheimer-Snyder collapse \cite{oppenheimer&snyder39}---all the shells become singular at the same time, and thus none of the shells cross \cite{newtonian}. For inhomogeneous matter configurations, however, the proper time for collapse depends on the (comoving) coordinate radius $r$, and thus the piling up of neighboring matter shells at finite proper radius can occur, thereby producing two-dimensional caustics where the energy density and some curvature components diverge \cite{yodzis&seifert&hagen73}. These singularities can be locally naked, but they are gravitationally weak \cite{newman86,nolan00}---curvature invariants and tidal forces remain finite---and, from the viewpoint of geodesic completeness, analytic continuations of the metric can always be found, in a distributional sense, in the neighborhood of the singularity \cite{papapetrou&hamoui67}. In this respect, shell crossings are not genuine physical singularities; rather, they signal the intersection of matter flow lines at a certain spacelike surface (in spherically symmetric geometries), and thus the breakdown of the model beyond that surface. (It is worth noting that they also occur in spherical inhomogeneous Newtonian gravitational collapse).

The inclusion of an electric charge density in spherical dust models has been considered by various authors \cite{novikov67,vickers73,ori90}. The principal physical motivation for such a generalization was the possibility of a ``gravitational bounce'' at the late stages of collapse, that could prevent the formation of a black hole \cite{novikov67}. However, a detailed analysis by Ori \cite{ori91} showed that shell crossings are inevitable in the collapse of weakly (the absolute value of the specific charge is less than unity) charged spherical dust shells. Inner shells are gravitationally more weakly bound than outer shells---due to the larger Lorentz repulsion and weaker gravitational potential---and thus collapse slower than outer shells, thereby leading to shell crossings {\em before} the (otherwise) minimum bounce radius is reached.

One might naively hope that the introduction of a positive cosmological constant, $\Lambda$, could prevent the occurrence of shell crossing---thus allowing for singularity avoidance---by increasing the outward acceleration of outer shells with respect to that of inner shells, i.e., by making outer shells more weakly bound (since the effective `gravitational potential' contains a repulsive $\Lambda r^{2}$ term that dominates at large radii) relative to inner ones.

In this paper, we generalize the existing results for shell crossing in weakly charged dust collapse, to include the contribution of a positive cosmological constant. Our motivation is fourfold: (i) Recent observations of high-redshift Type Ia supernovae \cite{riess98,perlmutter98} and peculiar motion of low-redshift galaxies \cite{zehani&dekel99}, appear to indicate that the present radius of the universe is accelerating, thus suggesting the existence of a positive cosmological constant, $\Lambda>0$; (ii) such a non-vanishing cosmological constant qualitatively changes the standard asymptotically flat picture of charged dust collapse, allowing, in particular, for an altogether different causal structure in the static limit: the Reisser-Nordstr\" om-de Sitter (RNdS) spacetime \cite{mellor&moss90}; (iii) the final exterior geometry of spherically charged dust configurations is similar to that of (more realistic) neutral rotating configurations \cite{lopez95}; (iv) the inclusion of a positive cosmological constant could, conceivably, prevent the occurrence of shell crossing, thereby allowing---at least in principle---for a singularity-free `bounce' model.

For definiteness, we shall refer to spherically symmetric solutions of the Einstein equations with charged dust and a cosmological constant, as generalized Tolman-Bondi metrics.

Contrary to the naive expectation, we find that the inclusion of a positive cosmological constant does {\em not} prevent the occurrence of shell crossing: the latter always occurs, irrespective of how large (but finite) the former is, for time-symmetric initial data. For non time-symmetric initial data, the occurrence of shell crossing is determined by the strength of the initial data; for sufficiently steep density gradients and non-negative initial velocity gradients, a non-zero-measure set of shells will cross at finite area radius near the center.

This paper is organized as follows: Section II derives the Tolman-Bondi family of metrics from the Einstein equations and discusses shell crossings in such spacetimes. In Sec. III, the spherical collapse of weakly charged dust with a cosmological constant is studied; the Einstein-Maxwell equations are reduced to a coupled set of first-order PDE's and an energy-type equation is obtained. Section IV gives a proof of the inevitability of shell crossing, near the center, based on a detailed analysis of the shape of the effective potential in the energy equation. In Sec. V, we  give an alternative proof of the occurrence of shell crossing in generalized Tolman-Bondi spacetimes, based on the point-particle/dust-shell analogy ({\em free surface} approximation). Section VI contains an analysis of the uncharged case, with $\Lambda>0$. Section VII concludes with a discussion and summary.

Natural geometrized units, in which $G=c=1$, are used throughout.

\section{TOLMAN-BONDI SPACETIMES}

For completeness, we present here a brief description of Tolman-Bondi metrics and shell crossings in neutral dust collapse in asymptotically flat spacetimes.

The Tolman-Bondi family of solutions is given by a spherically symmetric metric, written here in normal Gaussian coordinates $\{t,r,\theta,\phi\}$:
\bqa
ds^{2}&=&-dt^{2}+e^{-2\Psi(t,r)}dr^{2}+R^{2}(t,r)d\Omega^{2}, \label{metb}\\
d\Omega&\equiv&d\theta^{2}+\sin^{2}\theta d\phi^{2},
\eqa
together with the stress-energy tensor for dust:
\beq
T_{ab}=\rho(t,r)u_{a}u_{b}=\rho\delta_{a}^{t}\delta_{b}^{t},
\eeq
where $u^{a}=\delta^{a}_{t}$ is the 4-velocity of a dust element and $\rho(t,r)$ the energy density.

With the metric (\ref{metb}), the independent non-vanishing Einstein tensor components are
\bqa
G_{tt}&=&R^{-2}[-Re^{2\Psi}(2R'\Psi'+2R''+R^{-1}R'^{2}) \nonumber \\
&&-2\dot{R}\dot{\Psi}R+1+\dot{R}^{2}], \\
G_{rt}&=&-2R^{-1}(\dot{R}'+R'\dot{\Psi}), \\
G_{rr}&=&-R^{-2}\left[e^{-2\Psi}(2\ddot{R}R+\dot{R}^{2}+1)-R'^{2}\right], \\
G_{\theta\theta}&=&\sin^{-2}\theta\,G_{\phi\phi}=R(\dot{R}\dot{\Psi}+R'\Psi'e^{2\Psi} \nonumber \\
&&+R''e^{2\Psi}-\ddot{R}+\ddot{\Psi}R-\dot{\Psi}^{2}R),
\eqa
where the overdot and prime denote partial differentiation with respect to $t$ and $r$, respectively.

Introducing the auxiliary functions
\bqa
k(t,r)&\equiv&1-e^{2\Psi}R'^{2}, \label{kappa} \\
m(t,r)&\equiv&\fr{1}{2}R(\dot{R}^{2}+k), \label{masstb}
\eqa
Einstein's equations\footnote{Since there are only three functions to be determined and four equations, only three of these are independent, with the remaining one acting as a constraint. We take Eqs. (\ref{tb1})-(\ref{tb3}) as our complete set, and Eq. (\ref{mtb}) as the constraint equation, since it provides a simple relation between the initial data and the initial mass profile.} simplify greatly to
\bqa
\dot{R}^{2}&=&2mR^{-1}-k, \label{tb1} \\
\dot{k}&=&0, \label{tb2} \\
\dot{m}&=&0, \label{tb3}
\eqa
with the constraint
\beq
m'=4\pi R^{2}R'T_{tt}. \label{mtb}
\eeq

The model is then reduced to a single first-order PDE for the area radius $R(t,r)$, given by Eq. (\ref{tb1}) with
\beq
m(r)=4\pi\int_{0}^{r} R^{2}(0,\tilde{r})R'(0,\tilde{r})\rho(0,\tilde{r})d\tilde{r}.
\eeq
If $m(r)$ tends to a constant at spatial infinity, then $M=\lim_{r\rightarrow+\infty} m(r)$ is the ADM mass of the spacetime.

The metric (\ref{metb}) with $e^{2\Psi}=[1-k(r)]/R'^{2}$, together with Eqs. (\ref{tb1}) and (\ref{mtb}), fully determine the Tolman-Bondi family of solutions. Included in this family are the Schwarzschild metric ($m=\mbox{const.}$), the Einstein-de Sitter universe ($R\propto rt^{2/3}$, $k=0$), and the closed Friedmann universe [$R=ra(t)$, $k=r^{2}$].

The general\footnote{The $k>0$ solution corresponds to gravitationally bound configurations, and the $k=0$ case---discussed below---to marginally bound systems. For the unbound case, $k<0$, an analytical solution has also been obtained in closed form (see e.g. \cite{humphreys&maartens&matravers98} for a systematic treatment of spherical dust spacetimes), but it is of little interest to the implosion situation we are interested in.} ($k>0$) Tolman-Bondi solution can be easily obtained by parametric integration of Eq. (\ref{tb1}):
\bqa
t(\eta,r)&=&t_{0}(r)+mk^{-\fr{3}{2}}(\eta+\sin\eta), \label{t1} \\
R(\eta,r)&=&\fr{2m}{k}\cos^{2}\left(\fr{\eta}{2}\right), \label{R1}
\eqa
where $0\leq\eta\leq\pi$, and $t_{0}(r)$ is an arbitrary constant of integration to be fixed by the initial data, $\dot{R}(0,r)\equiv v(r)$, via
\beq
t_{0}(r)=mv^{-3}\tan^{3}\left(\fr{\eta_{0}(r)}{2}\right)\left[\eta_{0}(r)+\sin\eta_{0}(r)\right].
\eeq
For time-symmetric initial data, $v(r)=t_{0}(r)=0$, which implies $k(r)=2m/r$.

The radial coordinate $r$ is merely a label for the different shells, and we can therefore fix the radial coordinate gauge by equating the initial area radius to the coordinate radius:
\beq
R(0,r)=r,
\eeq
so that Eqs. (\ref{t1}) and (\ref{R1}) simplify to
\bqa
t(\eta,r)&=&\left(\fr{r^{3}}{8m}\right)^{\fr{1}{2}}(\eta+\sin\eta), \\
R(\eta,r)&=&r\cos^{2}\fr{\eta}{2}.
\eqa
A shell with initial proper area $4\pi r^{2}$ will thus collapse to vanishing area radius in a (comoving) time
\beq
t_{\rm coll}(r)=\pi\sqrt{\fr{r^{3}}{8m}}.
\eeq
For inhomogeneous mass distributions ($m\neq\mbox{const.}\times r^{3}$), different shells become singular at different times; in the homogeneous case, all the shells collapse to zero area radius at the same time \cite{newtonian}. 

The relevant derivatives of the area radius are
\bqa
R'&=&\cos^{2}\left(\fr{\eta}{2}\right)+\fr{r}{4}\left(\fr{3}{r}-\fr{m'}{m}\right)(\eta+\sin\eta)\tan\left(\fr{\eta}{2}\right), \label{Rpr} \\
\dot{R}&=&-\sqrt{\fr{2m}{r}}\tan\left(\fr{\eta}{2}\right).
\eqa

\subsection{Shell crossing in Tolman-Bondi spacetimes}

In the context of Tolman-Bondi metrics, shell crossings are defined by the loci of events given by
\beq
R'=0\;\;\;\mbox{and}\;\;\;R>0. \label{scs}
\eeq
At $R=R'=0$, a shell {\em focusing} singularity is said to occur. Unlike shell crossings, this central shell focusing singularity does not admit any metric extension through it and the spacetime is therefore geodesically incomplete (see e.g. \cite{joshi&dwivedi93}). It has been shown that shell focusing singularities can be naked \cite{lake92,lemos92} and gravitationally strong (finite physical volumes are `crushed' to zero at the singularity; see \cite{singh96} and references therein), although massless \cite{dwivedi&joshi92}. For the remainder of this paper we shall be concerned only with shell crossing singularities, as defined by (\ref{scs}).

Clearly, the necessary and sufficient condition for shell crossings is
\beq
t'_{\rm coll}<0,
\eeq
which is a condition on the initial mass distribution:
\beq
\fr{m'}{m}>\fr{3}{r}.
\eeq
Only sufficiently steep density distributions can provide a large enough specific binding energy, $k(r)$, that allows outer shells to overlap inner ones. This condition is the same as that obtained by requiring $R'=0$, with $R'$ given by Eq. (\ref{Rpr}).

\subsubsection{Marginally bound configurations}

The special case of $k(r)=0$ corresponds to a marginally bound matter configuration. It is the simplest case, and all the relevant expressions can be obtained analytically in closed form.

The dynamical equation
\beq
\dot{R}^{2}=\fr{2m}{R},
\eeq
is trivially integrated to
\beq
R(t,r)=r\left[1-\fr{t}{t_{\rm c}(r)}\right]^{\fr{2}{3}},
\eeq
where
\beq
t_{\rm c}(r)=\sqrt{\fr{2r^{3}}{9m}},
\eeq
is the proper time for the complete collapse of a spherical shell with initial area radius $r$.
Shell crossings occur when
\bqa
R'&=&\left(1-\fr{t}{t_{\rm c}}\right)^{-\fr{1}{3}}\left[1-\fr{t}{t_{\rm c}}+t_{\rm c}\gamma(r)\right]=0, \\
\gamma(r)&\equiv&1-\fr{r}{3}\fr{m'}{m}.
\eqa
$R'$ vanishes at
\beq
t_{\rm s}=t_{\rm c}(1+t_{\rm c}\gamma).
\eeq
Hence, the necessary and sufficient condition for shell crossing is
\beq
t_{\rm s}<t_{\rm c},
\eeq
which is exactly the same as the one for the time-symmetric $k>0$ case, $m'/m>3/r$. [Note, however, that the $k=0$ case corresponds to non-time-symmetric initial data, $\dot{R}(0,r)=\pm\sqrt{2m/r}$].

\section{GENERALIZED TOLMAN-BONDI SPACETIMES}

We consider a generalization of the original Tolman-Bondi metrics, that includes a charge density distribution $\mu(t,r)$ and a cosmological constant, $\Lambda$. As pointed out by Ori \cite{ori90}, the existence of a non-vanishing charge density implies that the charged matter shells do not follow geodesic motion, and thus their comoving time $t$ no longer equals proper time $\tau$, which precludes the use of Gaussian normal coordinates ($g_{tt}=-1$). We therefore consider a general spherically symmetric metric in the standard spherical coordinates $\{t,r,\theta,\phi\}$:
\beq
ds^{2}=-e^{2\Phi(t,r)}dt^{2}+e^{-2\Psi(t,r)}dr^{2}+R^{2}(t,r)d\Omega^{2}, \label{met}
\eeq
where $R(t,r)$ is the proper area radius and $d\Omega^{2}=d\theta^{2}+\sin^{2}\theta d\phi^{2}$ is the canonical metric of the unit two-sphere.

The stress-energy tensor is that of a charged dust fluid with a cosmological constant, $\Lambda$:
\beq
T_{ab}=\fr{1}{4\pi}(F_{ac}F_{b}^{\;c}-\fr{1}{4}g_{ab}F_{de}F^{de})+\rho u_{a}u_{b}-\fr{\Lambda}{8\pi}g_{ab},
\eeq
where $F_{ab}$ is the Maxwell tensor, $\rho$ is the energy density, and $u^{a}=e^{-\Phi}\delta_{t}^{a}$ is the four-velocity of a charged dust element. $F_{ab}$ is constrained by the Maxwell equations
\bqa
\nabla_{b}F^{ab}&=&-4\pi j^{a}, \\
\nabla_{[a}F_{bc]}&=&0,
\eqa
where the four-current is $j^{a}=\mu u^{a}$, and $\mu$ is the charge density. Spherical symmetry implies that the only non-vanishing component of the Maxwell tensor can be chosen to be
\beq
F_{rt}=e^{\Phi-\Psi}\fr{Q}{R^{2}},
\eeq
where $Q(t,r)$ is the total charge inside a shell $\Sigma_{t}$ with proper area $4\pi R^{2}(t,r)$:
\bqa
Q(t,r)&=&\int_{\Sigma_{t}} j^{a}d\Sigma_{a}=\int_{0}^{2\pi}\int_{0}^{\pi}\int_{0}^{r} \sqrt{-g}j^{t}dr'd\theta d\phi \nonumber \\
&=&4\pi\int_{0}^{r} \mu e^{-\Psi}R^{2}d\bar{r}. \label{charg}
\eqa

With the metric (\ref{met}) the non-vanishing Einstein tensor components are:
\bqa
G_{tt}&=&\fr{e^{2(\Phi+\Psi)}}{R^{2}}\left[R'^{2}+2R(R''+\Psi'R')\right] \nonumber \\
&&-\fr{\dot{R}}{R^{2}}(\dot{R}+2R\dot{\Psi}), \\
G_{tr}&=&\fr{2}{R}\left(\dot{R}'-\dot{R}\Phi'+\dot{\Psi}R'\right), \\
G_{rr}&=&\fr{e^{-2(\Psi+\Phi)}}{R^{2}}\left[2R\ddot{R}+\dot{R}(\dot{R}-2R\dot{\Phi})\right] \nonumber \\
&&-\fr{R'}{R^{2}}(R'+2R\Phi'), \\
G_{\theta\theta}&=&-R\{e^{-2\Phi}\left[-\ddot{R}+\dot{R}(\dot{\Phi}+\dot{\Psi})+R(\ddot{\Psi}-\dot{\Psi}^{2}-\dot{\Psi}\dot{\Phi})\right] \nonumber \\ 
&&\!\!\!\!\!\!\!\!\!\!\!\!\!\!\!\!\!\!\!+e^{2\Psi}\left[R''+R'(\Phi'+\Psi')+R(\Phi''+\Phi'^{2}+\Phi'\Psi')\right]\}, \\
G_{\phi\phi}&=&\sin^{2}\theta G_{\theta\theta}.
\eqa

The non-vanishing components of the stress-energy tensor are
\bqa
T_{tt}&=&\fr{e^{2\Phi}}{8\pi}\left(\fr{Q^{2}}{R^{4}}+\Lambda+8\pi\rho\right), \\
T_{rr}&=&\fr{e^{-2\Psi}}{8\pi}\left(-\Lambda-\fr{Q^{2}}{R^{4}}\right), \\
T_{\theta\theta}&=&\sin^{-2}\theta T_{\phi\phi}=\fr{R^{2}}{8\pi}\left(-\Lambda+\fr{Q^{2}}{R^{4}}\right).
\eqa

Introducing the auxiliary functions
\bqa
k(t,r)&\equiv&1-e^{2\Psi}R'^{2}, \label{kay} \\
m(t,r)&\equiv&\fr{1}{2}R\left(e^{-2\Phi}\dot{R}^{2}+k+\fr{Q^{2}}{R^{2}}-\fr{\Lambda}{3}R^{2}\right), \label{mas}
\eqa
the Einstein-Maxwell equations can be recast as
\bqa
e^{-2\Phi}\dot{R}^{2}&=&-Q^{2}R^{-2}+2mR^{-1}-k+\fr{\Lambda}{3}R^{2}, \label{e1} \\
\dot{k}&=&2\fr{\dot{R}}{R'}(k-1)\Phi', \label{e2} \\
\dot{m}&=&0, \label{e3} \\
\dot{Q}&=&0, \label{e3b} \\
m'&=&4\pi R^{2}R'\rho+QQ'R^{-1}+\fr{\Lambda}{2}R^{2}R', \label{e4} \\
Q'&=&4\pi R^{2}e^{-\Psi}\mu. \label{e5}
\eqa
Equation (\ref{e1}) is just Eq. (\ref{mas}); Eqs. (\ref{e2})-(\ref{e4}) follow directly from the Einstein equations, and Eq. (\ref{e5}) follows from Eq. (\ref{charg}). The equation for $k'$ is algebraically complicated and it proves more convenient to use an equivalent equation given by the local conservation of energy-momentum, $\nabla_{a}T^{a}_{b}=0$ (where we took $b\equiv r$):
\beq
QQ'=4\pi\rho R^{4}\Phi'. \label{e6}
\eeq
This equation expresses the balance between Lorentz and gravitational `forces' in the comoving frame. The other equation ($b\equiv t$) from the local conservation of energy-momentum gives the conservation of specific charge:
\beq
\partial_{t} \left(\fr{\mu}{\rho}\right)\equiv\dot{\epsilon}=0.
\eeq
Equations (\ref{kay}), (\ref{e1}), (\ref{e2}), and (\ref{e6}) form a complete set; the remaining independent equations are constraints. We have thus reduced the Einstein-Maxwell system to a set of coupled first-order PDE's. It is not the purpose of the present paper to obtain an explicit general solution in closed form---this has been done, in a different coordinate system, by Ori \cite{ori90}. As we shall see shortly, first integrals of the equations of motion suffice to prove the inevitability of shell crossings in generalized Tolman-Bondi models\footnote{This is essentially due to the fact that shell crossings are {\em dynamical} processes, whence the knowledge of the first `time' derivative of the spheres' radii suffices to analyze their {\em relative} motion.}.

Using Eqs. (\ref{kay}), (\ref{e5}), and (\ref{e6}), Eq. (\ref{e2}) integrates to
\beq
k(R,r)=1-\left[\epsilon\fr{Q}{R}+W(r)\right]^{2}, \label{dbu}
\eeq
where $W(r)$ is a free function to be fixed by the initial data, via Eq. (\ref{e1}) evaluated at $t=0$. Comparison with the neutral case, leads to the interpretation of $W$ as a measure of the binding energy of the system per unit mass. For $\Lambda=0$, the configuration is gravitationally unbound, marginally bound, or bound, when $W>1$, $W=1$, or $W>1$, respectively. 

For time-symmetric initial data, $\dot{R}(0,r)=0$, with the scaling $R(0,r)=r$, we have then
\beq
W(r)=\epsilon\fr{Q}{r}+\sqrt{1-\fr{2m}{r}+\fr{Q^{2}}{r^{2}}-\fr{\Lambda}{3}r^{2}}, \label{energy}
\eeq
where the `$+$' sign was fixed by consistency with the asymptotic $\Lambda=Q=0$ limit.

Regularity at the origin requires \cite{vickers73}:
\bqa
\lim_{r\rightarrow0^{+}} \fr{Q}{r}&=&0, \\
\lim_{r\rightarrow0^{+}} \fr{m}{r}&=&0, \\
\lim_{r\rightarrow0^{+}} |\epsilon|&=&|\epsilon_{0}| \in (0,1),
\eqa
which implies
\beq
W(0)\equiv W_{0}=1.
\eeq

Summarizing, the only non-trivial dynamical equation is
\beq
e^{-2\Phi}\dot{R}^{2}+U(R,r)=0, \label{gtbdyn}
\eeq
with the effective potential
\bqa
U(R,r)&=&a_{0}+a_{1}R^{-1}+a_{2}R^{-2}+a_{3}R^{2}, \\
a_{0}(r)&=&1-W^{2}, \\
a_{1}(r)&=&2m\left(\epsilon\fr{Q}{m}W-1\right), \\
a_{2}(r)&=&Q^{2}(1-\epsilon^{2}), \\
a_{3}&=&-\fr{\Lambda}{3}.
\eqa

\section{EFFECTIVE POTENTIAL ANALYSIS}

In this section, we prove two Propositions that show that shell crossing is inevitable in (weakly charged) generalized Tolman-Bondi spacetimes.

{\bf Proposition 1.} For weakly charged initial data, there exists $r_{*}>0$, such that the effective potential $U(R,r)$ has, for $r\in(0,r_{*})$, three distinct real positive zeros, $R_{1}<R_{2}<R_{3}$, satisfying (see Fig. 1)
\bqa
U(R_{1}<R<R_{2},r)<0, \label{uc1} \\
U(R_{2}<R<R_{3},r)>0. \label{uc2}
\eqa

Sketch of the proof. We prove the above proposition by showing that for $0<r<r_{*}\ll1$, $U(R,r)$ has a local negative minimum at $R_{\rm m}>0$ and a local positive maximum at $R_{\rm M}>R_{\rm m}$. Since $\lim_{R\rightarrow0^{+}} U=+\infty$ and $\lim_{R\rightarrow+\infty} U=-\infty$, it follows that $U(R,r)$ has three distinct real positive zeros, obeying conditions (\ref{uc1})-(\ref{uc2}).

{\bf Proof.} Let us rewrite
\bqa
U(R,r)&=&a_{3}R^{-2}H(R,r), \\
H(R,r)&=&R^{4}+c_{2}R^{2}+c_{1}R+c_{0}, \label{bigh} \\
c_{0}&\equiv&-\fr{3}{\Lambda}Q^{2}(1-\epsilon^{2}), \label{c0} \\
c_{1}&\equiv&-\fr{6}{\Lambda}m\left(\epsilon W\fr{Q}{m}-1\right), \label{c1} \\
c_{2}&\equiv&-\fr{3}{\Lambda}(1-W^{2}). \label{c2}
\eqa
The relevant limits are now
\bqa
\lim_{R\rightarrow0^{+}} H&=&c_{0}<0, \\
\lim_{R\rightarrow+\infty} H&=&+\infty.
\eqa
Although the local extrema of $U$ are different from those of $H$, all (non-zero) roots of $H$ are also roots of $U$, and we can thus study the qualitative behavior of $U$ by analyzing $H$ and its first partial derivative with respect to $R$:
\beq
\fr{1}{4}\fr{\partial H}{\partial R}=R^{3}+\fr{1}{2}c_{2}R+\fr{1}{2}c_{1}.
\eeq
Now, $\partial_{R}H$ has three distinct real zeros provided \cite{abramovitz&stegun64}
\beq
\left(\fr{W^{2}-1}{2}\right)^{3}<-\Lambda\left[6m\left(\epsilon W\fr{Q}{m}-1\right)\right]^{2}. \label{cond1}
\eeq
Since the right-hand-side of the inequality is manifestly negative, this requires $W<1$. Taylor expanding $Q(r)$ and $m(r)$ near the origin, using Eqs. (\ref{e4})-(\ref{e6}), and Eq. (\ref{dbu}) evaluated at $t=0$, with the scaling $R(0,r)=r$, we have
\beq
W(r)=1-\left[\fr{\Lambda}{3}+\fr{4\pi}{3}\rho_{0}(1-\epsilon_{0}^{2})\right]r^{2} +{\mathcal{O}}(r^{4}),
\eeq
for $r\in(0,\xi_{1})$, where $\xi_{1}\ll1$. Condition (\ref{cond1}) is trivially satisfied for $0<r<\xi_{1}$.

We will now show that of the three real roots of $\partial_{R}H$, two are positive, say $\tilde{R}_{1}$ and $\tilde{R}_{2}$, such that $H(\tilde{R}_{1})H(\tilde{R}_{2})<0$, thereby guaranteeing that $H$ has a local positive maximum at $\tilde{R}_{1}\in(R_{1},R_{2})$ and a local negative minimum at $\tilde{R}_{2}\in(R_{2},R_{3})$.

The roots $\tilde{R}_{i}$ ($i=1,2,3$) of the polynomial $\partial_{R}H$ satisfy \cite{abramovitz&stegun64}:
\bqa
\sum_{i} \tilde{R}_{i}&=&0, \label{p1} \\
\sum_{i\neq i} \tilde{R}_{i}\tilde{R}_{j}&=&\fr{3}{2\Lambda}(W^{2}-1), \label{p2} \\
\prod \tilde{R}_{i}&=&\fr{3m}{2\Lambda}\left(\epsilon W\fr{Q}{m}-1\right), \label{p3}
\eqa
and can be explicitly given as
\bqa
\tilde{R}_{1}&=&2\sqrt{\chi}\cos\left(\fr{\Theta}{3}\right), \label{rt1} \\
\tilde{R}_{2}&=&2\sqrt{\chi}\cos\left(\fr{\Theta}{3}+\fr{2\pi}{3}\right), \label{rt2} \\
\tilde{R}_{3}&=&2\sqrt{\chi}\cos\left(\fr{\Theta}{3}+\fr{4\pi}{3}\right), \label{rt3}
\eqa
where
\bqa
\chi&=&\fr{1-W^{2}}{2\Lambda}, \\
\cos\Theta&=&3\sqrt{\fr{\Lambda}{2}}\fr{\epsilon QW-m}{(1-W^{2})^{3/2}}.
\eqa
Since $\cos$ is anti-periodic in $\pi$ and the trigonometric arguments in Eqs. (\ref{rt1})-(\ref{rt2}) are separated by a $2\pi/3$ phase, it suffices to require that two of the roots are positive, the third one being then necessarily negative. We therefore impose $\pi/2<\Theta<3\pi/2$ (thereby guaranteeing that $\tilde{R}_{1}$ and $\tilde{R}_{2}$ are positive), which translates to
\beq
3m\left(1-\epsilon W\fr{Q}{m}\right)<\sqrt{\fr{2}{\Lambda}}(1-W^{2})^{\fr{3}{2}}. \label{cond2}
\eeq
Near the origin, for $0<r<\xi_{2}\ll1$, this becomes
\beq
\fr{4}{\sqrt{\Lambda}}\left[\fr{4\pi}{3}\rho_{0}(1-\epsilon_{0}^{2})+\fr{\Lambda}{3}\right]^{\fr{3}{2}}r^{\fr{3}{2}}+{\mathcal{O}}(r^{3})>0,
\eeq
which is trivially satisfied; thus $\tilde{R}_{1}$ and $\tilde{R}_{2}$ are positive, for $r\in(0,\xi_{2})$.

Now, it remains to show that $H(\tilde{R}_{1})H(\tilde{R}_{2})<0$. From Eq. (\ref{bigh}) together with properties (\ref{p1})-(\ref{p3}), and setting $\tilde{R}_{3}=-x$, we obtain
\bqa
H(\tilde{R}_{1})H(\tilde{R}_{2})&\equiv&f(x)=-\fr{1}{4}\left(\fr{c_{2}}{4}\right)^{2}c_{1}^{2}x^{-2} \nonumber \\
&&-\left[\fr{1}{4}c_{0}c_{1}c_{2}+\left(\fr{c_{1}}{4}\right)^{3}\right]x^{-1}-\fr{3}{4}c_{0}c_{1}x \nonumber \\
&&+c_{0}^{2}+\fr{3}{2}\left(\fr{c_{1}}{4}\right)^{2}c_{2}.
\eqa
Using Eqs. (\ref{c0})-(\ref{c2}), it is straightforward to check that $f$ is a monotonically increasing function of $x$. From Eq. (\ref{rt3}) we have $x\leq\sqrt{-c_{2}/6}\equiv x_{\rm M}$. Thus, it suffices to show that $f(x_{\rm M})<0$:
\beq
c_{0}^{2}\left[1-\fr{3}{4}c_{1}\sqrt{\fr{-c_{2}}{6}}\left(1-\fr{\sqrt{2/3}}{c_{0}}\right)\right]-\left(\fr{c_{1}}{4}\right)^{3}\sqrt{\fr{6}{-c_{2}}}<0. \label{cond3}
\eeq
Near the origin, for $r\in(0,\xi_{3})$, with $\xi_{3}\ll1$, we have
\beq
-\sqrt{\fr{6}{-\alpha_{2}}}\left(\fr{\alpha_{1}}{4}\right)^{3}r^{8} +{\mathcal{O}}(r^{10}), \label{cd3} \\
\eeq
where
\bqa
\alpha_{1}&=&\fr{6}{\Lambda}\left(\fr{4\pi}{3}\rho_{0}+\fr{\Lambda}{3}\right)(1-\epsilon_{0}^{2}), \\
\alpha_{2}&=&-\fr{3}{\Lambda}\left[\fr{4\pi}{3}\rho_{0}(1-\epsilon_{0}^{2})+\fr{\Lambda}{3}\right].
\eqa
Condition (\ref{cond3}) is manifestly satisfied; hence $H(\tilde{R}_{1})H(\tilde{R}_{2})<0$, for $0<r<\xi_{3}$.

Now, let us take $r_{*}=\mbox{min}\,\{\xi_{1},\xi_{2},\xi_{3}\}$, such that for $r\in(0,r_{*})$ conditions (\ref{cond1}), (\ref{cond2}), and (\ref{cond3}) are mutually satisfied.

Summarizing, we have shown that there exists $r_{*}>0$, such that for $r\in(0,r_{*})$, $\partial_{R}H$ has two distinct real positive zeros, corresponding to the local extrema of $H(R,r)$. We have further shown that the product of the local extrema of $H$ is negative, which implies that $H$ has three distinct real positive roots, since $\lim_{R\rightarrow0^{+}} H=c_{0}<0$ and $\lim_{R\rightarrow+\infty} H=+\infty$. By construction, these are the {\em same} zeros of $U(R,r)$. Since $U(R,r)=a_{3}(r)R^{-2}H(R,r)$ and $a_{3}<0$, it follows that $U(R,r)$ is negative for $R\in(R_{1},R_{2})$ and positive for $R\in(R_{2},R_{3})$, as desired. $\blacksquare$

{\bf Proposition 2.} Consider a general solution of Eq. (\ref{e1}), with parameters $\{Q,m,\Lambda\}$ satisfying Proposition 1, and the spacetime metric determined by that solution. In such a spacetime, shell crossing always occurs from regular time-symmetric initial data, for a non-zero-measure set of spherical weakly charged matter shells.

{\bf Proof.} For $r>0$, $\lim_{R\rightarrow0^{+}} U(R,r)=+\infty$ and $\lim_{R\rightarrow+\infty} U(R,r)=-\infty$. Since $U$ is $C^{\infty}$ with respect to $R$, this implies---from Eq. (\ref{e1})---that there is a time $t_{\rm m}>0$ such that
\beq
\dot{R}(t_{\rm m},r)=0. \label{rdt0}
\eeq
Let us then define $R_{\rm m}(r)\equiv R(t_{\rm m},r)$.

Taking the total derivative of Eq. (\ref{e1}) with respect to $r$, we obtain
\beq
2e^{-2\Phi}\dot{R}\dot{R}'-2\Phi'e^{-2\Phi}\dot{R}^{2}+\fr{\partial U}{\partial R}R'+U'=0. \label{rdts}
\eeq
For time-symmetric initial data, $\dot{R}(0,r)=0$, and thus, from Eqs. (\ref{rdt0})-(\ref{rdts}) we have
\bqa
R'(0,r)&=&-U'(r,r)/\left(\fr{\partial U}{\partial R}\right)_{R=r}, \label{Rpr0} \\
R'(t_{\rm m},r)&=&-U'(R_{\rm m},r)/\left(\fr{\partial U}{\partial R}\right)_{R=R_{\rm m}}. \label{Rprs}
\eqa
We now examine the signs of the numerators in the equations above. From Eq. (\ref{gtbdyn}) we have
\bqa
U'(r,r)&=&a'_{0}+a'_{1}r^{-1}+a'_{2}r^{-2}, \\
U'(R_{\rm m},r)&=& a'_{0}+a'_{1}R^{-1}_{\rm m}+a'_{2}R^{-2}_{\rm m}.
\eqa
Setting $R(0,r)=r=R_{2}$ and $R_{\rm m}(r)=R_{1}$, it follows from Proposition 1 that $R_{\rm m}(r)<r$, and we thus can set $R_{\rm m}=\beta_{r}r$ for any given shell $r$, where $\beta_{r}<1$ depends on $r$. Near the origin, for $0<r<r_{\bigstar}\ll1$, we have then
\bqa
U'(r,r)&=&-\left[\fr{8\pi}{3}\rho_{0}(1-\epsilon_{0}^{2})+\Lambda\left(\fr{5}{3}-\epsilon^{2}_{0}\right)\right]r \nonumber \\
&&+{\mathcal{O}}(r^{3}), \\
U'(R_{\rm m},r)&=&-[\left(\fr{6}{\beta_{r}}-4\right)\fr{4\pi}{3}\rho_{0}(1-\epsilon_{0}^{2}) \nonumber \\
&&+\fr{\Lambda}{\beta_{r}}\left(1-\epsilon_{0}^{2}+\fr{2}{3}\beta_{r}\right)]r+{\mathcal{O}}(r^{3}).
\eqa
Both numerators are manifestly negative, for $r\in(0,r_{\bigstar})$.

Now, from Proposition 1, we have $(\partial U/\partial R)_{R=r}>0$ and $(\partial U/\partial R)_{R=R_{\rm m}}<0$. Since $U(0,r)=U(t_{\rm m},r)=0$, it follows from Eqs. (\ref{Rpr0})-(\ref{Rprs}) that
\beq
\fr{R'(0,r)}{R'(t_{\rm m},r)}<0.
\eeq
Hence, by continuity, there is a time $t_{\rm sc}\in(0,t_{\rm m})$, such that
\beq
R'(t_{\rm sc},r)=0, \; \forall \; r\in(0,\mbox{min}\{r_{*},r_{\bigstar}\}).
\eeq
This completes the proof. $\blacksquare$

\begin{figure}
\begin{center}
\epsfxsize=19pc
\epsffile{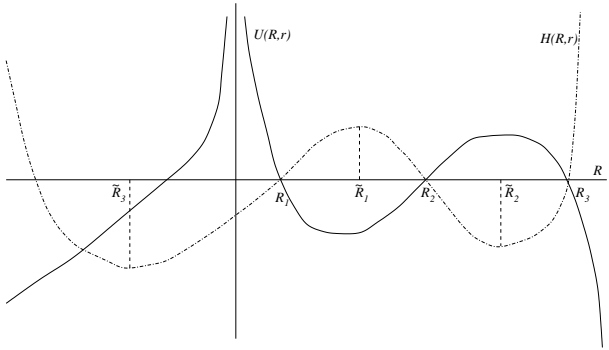}
\end{center}
\caption{Effective potential $U(R,r)$ and auxiliary polynomial $H(R,r)\propto R^{2}U(R,r)$, for $r\in(0,r^{*})$. Without shell crossing, a spherical shell with initial area radius $R(0,r)=r=R_{2}$ would collapse towards decreasing values of $R$ until reaching the minimum bounce radius $R_{\rm m}=R_{1}$ at some later time $t_{\rm m}$. However, shell crossing is {\em inevitable} at $t_{\rm sc}\in(0,t_{\rm m})$.\label{fig1}}
\end{figure}

\section{FREE SURFACE APPROACH ANALYSIS}

This Section gives an alternative proof of the inevitability of shell crossing in spherical weakly charged dust collapse with a cosmological constant. It relies on an immediate consequence of the generalized Birkhoff theorem \cite{MTW} (which can be extended to include a cosmological constant \cite{bonnor62}), the so-called {\em free surface} approach: as long as shells do not cross, their motion will depend solely on their total interior charge and mass, and thus each shell will move independently (albeit not geodesically) like a test particle in a RNdS background, with mass and charge equal to the total mass and charge of that shell (see e.g. \cite{ori91}).

\subsection{Particle motion in a Reisser-Nordstr\" om-de Sitter geometry}

In Schwarzschild coordinates $\{t,r,\theta,\phi\}$, the RNdS metric reads
\bqa
ds^{2}&=&-\Delta dt^{2}+\Delta^{-1}dr^{2}-r^{2}d\Omega^{2}, \label{rnds} \\ 
\Delta(r)&\equiv&1-\fr{2M}{r}+\fr{\bar{Q}^{2}}{r^{2}}-\fr{\Lambda}{3}r^{2},
\eqa
where $d\Omega^{2}=d\theta^{2}+\sin^{2}\theta d\phi^{2}$ is the canonical metric of the unit two-sphere.

The action for a particle of mass $\bar{m}$ and charge $q$ is
\beq
S[x^{a},e]=\int d\lambda\left(\fr{1}{2}e^{-1}\dot{x}^{a}\dot{x}^{b}g_{ab}-\fr{1}{2}\bar{m}^{2}e-q\dot{x}^{a}A_{a}\right), \label{action}
\eeq
where $\lambda$ is an affine parameter along the particle's worldline, $A_{a}=(\bar{Q}/r)\delta^{t}_{a}$ is the electromagnetic four-potential, and $e=e(\lambda)$ is the {\em einbein}, an independent function that generalizes the action for a point particle \cite{brink&vecchia&howe76}.

It is straighforward to show (cf. Appendix) that if
\beq
\left({\mathcal L}_{\xi}A\right)_{a}\equiv\xi^{b}A_{a,b}+A_{b}\xi^{b}_{,a}=0,
\eeq
for a Killing vector field $\xi^{a}$, then $S$ is invariant, to first-order in $\xi^{a}$, under the transformation $x^{a}\rightarrow x^{a}+\sigma\xi^{a}$, where $\sigma\in\Bbb{R}$. Denoting by $\pi^{a}=\bar{m}u^{a}-qA^{a}$ the generalized momentum of the particle, it then follows that the corresponding Noether charge, ${\mathcal E}=\xi^{a}\pi_{a}$, is a constant of motion:
\bqa
\fr{d{\mathcal E}}{d\lambda}&=&\left(\pi_{a}\nabla_{b}\xi^{a}+\xi^{a}\nabla_{b}\pi_{a}\right)u^{b}\nonumber \\
&=&\bar{m}u^{a}u^{b}\nabla_{b}\xi_{a}+q({\mathcal L}_{\xi}A_{a})u^{a}=0.
\eqa

One of the trivial Killing vectors of the RNdS metric is $\xi_{(t)}=\partial_{t}$. Thus, with the metric (\ref{rnds}), we have
\beq
E=\xi^{t}(\bar{m}u_{t}-qA_{t})=\bar{m}\Delta\fr{dt}{d\tau}+q\fr{\bar{Q}}{r}, \label{noether2}
\eeq
where $\tau$ is the particle's proper time. Since the geometry is not asymptotically flat, the constant of motion $E$ cannot be interpreted as the particle's total energy measured at spatial infinity; it should be regarded as an effective energy, that coincides with the particle's total energy at infinity in the $\Lambda\rightarrow0$ limit.

From Eq. (\ref{rnds}) we have, for a timelike radial worldline
\beq
\Delta \left(\fr{dt}{d\tau}\right)^{2}=1+\Delta^{-1}\left(\fr{dr}{d\tau}\right)^{2},
\eeq
which, together with Eq. (\ref{noether2}), gives
\bqa
&&\left(\fr{dr}{d\tau}\right)^{2}+V(r)=0, \label{radial} \\
&&V(r)=(1-\bar{W}^{2})-\left(1-\bar{W}\bar{\epsilon}\fr{\bar{Q}}{M}\right)\fr{2M}{r} \nonumber \\
&&+(1-\bar{\epsilon}^{2})\fr{\bar{Q}^{2}}{r^{2}}-\fr{\Lambda}{3}r^{2}, \label{vpt}
\eqa
where $\bar{W}\equiv E/\bar{m}$ is the particle's specific effective energy, and $\bar{\epsilon}\equiv q/\bar{m}$ its specific charge.

\subsection{Shell crossing analysis}

Equation (\ref{radial}), which governs the radial motion of test particles with specific charge $\epsilon=q/\bar{m}$ in a RNdS geometry with {\em fixed} total charge $\bar{Q}$ and mass $M$, will coincide with Eq. (\ref{gtbdyn}), which governs the radial motion of charged dust shells with total interior charge $Q(r)$ and mass $m(r)$, provided we make the following formal identifications [and set $\Phi=0$ in Eq. (\ref{gtbdyn}), thereby identifying proper and comoving times]:
\bqa
r&=&R, \\
\bar{W}&=&W, \\
M&=&m, \\
\bar{Q}&=&Q, \\
\bar{\epsilon}&=&\fr{Q}{m}.
\eqa
The potential $V(r)$ reads then
\bqa
&&V(r)=(1-W^{2})-\left(1-W\epsilon\fr{Q}{m}\right)\fr{2m}{r} \nonumber \\
&&+\left(1-\fr{Q^{2}}{m^{2}}\right)\fr{Q^{2}}{r^{2}}-\fr{\Lambda}{3}r^{2}. \label{vpt2}
\eqa
It governs the radial motion of a charged dust shell with proper area $4\pi r^{2}$ and specific charge $\epsilon(r)=Q(r)/m(r)$, in a RNdS background metric with total mass $m(r)$ and charge $Q(r)$. In the free surface approach, the only---but crucial---difference between shell and particle motion in a RNdS background, is that the latter travels on a fixed geometry, whereas the former travels on a geometry that is determined by the interior mass and charge of each shell.

Let us now consider two infinitesimally close shells, with wordlines $r_{0}(\tau)$ and $r_{1}(\tau)=r_{0}+\xi$ (where $0<\xi\ll1$), which are solutions of Eq. (\ref{radial}) with $V(r)<0$. As shown in Fig. 2, these two neighboring shells {\em will cross} provided
\beq
\left(\fr{dr}{d\tau}\right)_{r_{0}}>\left(\fr{dr}{d\tau}\right)_{r_{0}+\xi}.
\eeq
From Eq. (\ref{radial}), this translates into $V(r_{0}+\xi)<V(r_{0})$. Expanding $V(r_{0}+\xi)=V(r_{0})+V'(r_{0})\xi +{\mathcal O}(\xi^{2})$, to first order in $\xi$ we have then $V'(r_{0})\xi<0$. Since $\xi>0$, the condition for shell crossing is simply (dropping the `0' index)
\beq
V'(r)<0.
\eeq
Evaluating Eq. (\ref{vpt2}) near the origin, for $0<r<r_{\rm c}\ll1$, this condition becomes 
\beq
2\epsilon_{0}^{2}\fr{\Lambda}{3}\left(1+\fr{\Lambda}{8\pi\rho_{0}}\right)^{-1}r+{\mathcal O}(r^{3})>0,
\eeq
which is clearly satisfied for $r\in(0,r_{\rm c})$. Shell crossing is inevitable near the center.

\begin{figure}
\begin{center}
\epsfxsize=14pc
\epsffile{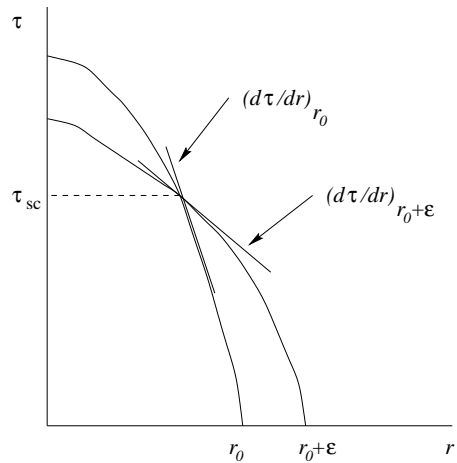}
\end{center}
\caption{Shell crossing for spherical weakly charged dust collapse in $\tau-r$ coordinates. For $r_{0}\in(0,r_{\rm c})$, any two neighboring shells, $r_{0}$ and $r_{0}+\xi$, will cross each other at $\tau=\tau_{\rm sc}$, when $V'(r_{0})<0$.\label{fig2}}
\end{figure}

\section{NEUTRAL CASE}

In this Section, we analyze separately the occurrence of shell crossings for neutral dust spheres in the presence of a positive cosmological constant, which is a simpler and more realistic model for macroscopic astrophysical objects.

Since there is no elecrostatic repulsion acting on the test particles, they move geodesically in a Schwarzschild-de Sitter geometry \cite{bonnor62}. We can therefore set $g_{tt}=-1$, thereby identifying proper and comoving times. The relevant dynamical equation is then
\bqa
\dot{R}^{2}+U(R,r)=0, \\
U(R,r)=1-W^{2}-\fr{2m}{R}-\fr{\Lambda}{3}R^{2}. \label{upt}
\eqa

We consider here the particular case of $W=1$, corresponding to a gravitationally unbound configuration, since it allows for an analytical solution in closed form:
\bqa
R(t,r)&=&\left(\fr{6m}{\Lambda}\right)^{\fr{1}{3}}\sinh^{2/3}\theta(t,r), \\
\theta(t,r)&\equiv&\fr{\sqrt{3\Lambda}}{2}\left[t_{\rm c}(r)-t\right],
\eqa
where $t_{\rm c}(r)$ is an arbitrary function to be fixed by the initial velocity profile via
\beq
\dot{R}(0,r)=-\sqrt{\fr{\Lambda}{3}}r\coth\left(\fr{\sqrt{3\Lambda}t_{\rm c}}{2}\right). \label{vel}
\eeq
The relevant derivatives of the area radius are
\bqa
R'(t,r)&=&R\left(\fr{m'}{3m}+\sqrt{\fr{\Lambda}{3}}t'_{\rm c}\coth\theta\right), \\
\dot{R}(t,r)&=&-\sqrt{\fr{\Lambda}{3}}R\coth\theta.
\eqa
Since $\theta\coth\theta>0$, the necessary and sufficient condition for shell crossings is
\beq
t'_{\rm c}<0.
\eeq
A trivial example is $t_{\rm c}=1/r$. For small $r$, $3m/m'\simeq r$, and $\coth\theta\simeq1/\theta$. Solving for $R'=0$ near the origin yields then
\beq
t_{\rm sc}(r)\simeq\fr{1}{3r}<t_{\rm c},
\eeq
thus confirming the occurrence of shell crossing near the center.

Since the initial velocity profile [cf. Eq. (\ref{vel})] is a monotonically increasing function of $t_{\rm c}(r)$, the condition $t'_{\rm c}<0$ is simply the requirement that outer shells have a sufficiently larger inward initial radial velocity than inner ones, to overcome the (comparatively larger, for outer shells) repulsive effect of $\Lambda$, thereby leading to shell crossing before complete collapse.

We note that since $W=1$ corresponds to unbound systems, the above criterion for shell crossing is likely to be relaxed for bound configurations. For $W<1$, for each shell $r$, there is a critical value $\Lambda_{\rm c}(t)$---which is the root of Eq. (\ref{upt}) for fixed $r$---such that the system is gravitationally bound for $\Lambda<\Lambda_{\rm c}$, and unbound for $\Lambda>\Lambda_{\rm c}$.

\section{DISCUSSION AND CONCLUSIONS}

We have shown that the inclusion of a positive cosmological constant in spherical charged dust collapse does {\em not} prevent the occurrence of shell crossing near the center. Heuristically, this can be explained by the fact that $\Lambda$ has a long range effect, whereas the relevant physics for shell crossing occurs near the center, where Lorentz and gravitational `forces' compete and the $\Lambda$ repulsion is negligible.

Although the free surface approach yields a rather simple method for proving the inevitability of shell crossing near the center, being a purely kinematical analysis it cannot relate the initial data to the motion of the shells. Reducing Einstein's equations to first integrals of motion enabled a more physical analysis, where the dynamics of collapse is determined by the choice of initial data. 

Proposition 1 showed that in a small neighborhood near the center of the matter distribution, the effective potential is such that a shell with initial area radius $r=R_{2}$ will collapse towards smaller values of $R$ and, provided there is no shell crossing, reach the minimum bounce radius $R_{1}$. Proposition 2 showed that for a such a shell [$R(0,r)=r=R_{2}$] shell crossing will {\em inevitably} occur {\em before} the minimum bounce radius is reached.

Proposition 2 used the assumption of time-symmetric initial data. This ansatz is by no means essential. For non-time-symmetric implosion situations, two possibilities arise: (i) the initial data is sufficiently strong, such that {\em all} shells have $\dot{R}(0,r)<0$, or (ii) a non-zero-measure set of shells has $\dot{R}(0,r)>0$, and therefore initially starts to expand towards increasing values of $R$, until each shell reaches a maximum area radius $R_{\rm max}(r)$ and then collapses back through its initial radius. Case (ii) reduces to the time-symmetric situation when $R(t,r)=R_{\rm max}$ and is thus covered by Proposition 2. For case (i), the occurrence---or lack thereof---of shell crossing is determined by the initial velocity profile, $v(r)\equiv \dot{R}(0,r)$ for a given mass $m(r)$ and charge $Q(r)$ [hence specific charge, $\epsilon(r)$] distribution. Clearly, if $v<0$ and $v'=0$ then shell crossing will occur, since it does when $v=v'=0$: all the shells are differentially accelerated in the same manner, irrespective of their initial velocity profile. It then follows that shell crossing will occur for any $v(r)<0$, provided $v'\leq0$. 

For the neutral case, shell crossing occurs---even for gravitationally unbound matter configurations---provided the initial velocity profile is sufficiently steep, irrespective of how large (but finite) $\Lambda$ may be. Unlike Lorentz `forces', which become more noticeable as collapse proceeds and the area radius of the shells decreases, the $\Lambda$ repulsion becomes increasingly unimportant at late times (i.e., in the strong-field region, at small radii), whence the criterion for shell crossing becomes analogous to that for uncharged dust collapse in an asymptotically flat spacetime.

\section*{ACKNOWLEDGMENTS}

I am grateful to Patrick Brady and Kip Thorne for useful discussions. This work was supported by F.C.T. (Portugal) Grant PRAXIS XXI-BPD-16301-98, and by NSF Grant AST-9731698.

\section*{APPENDIX}

Consider the following action:
\beq
S[x^{a},e]=\int d\lambda\left(\fr{1}{2}e^{-1}\dot{x}^{a}\dot{x}^{b}g_{ab}-\fr{1}{2}m^{2}e-q\dot{x}^{a}A_{a}\right),
\eeq
where $\lambda$ is an affine parameter along the particle's worldline, $A_{a}$ is the electromagnetic four-potential, and $e=e(\lambda)$ is the {\em einbein}. Assume that there is a Killing vector field $\xi^{a}$, such that $({\mathcal L}_{\xi}A)_{a}=0$. Under the transformation $x^{a}\rightarrow \bar{x}^{a}=x^{a}+\sigma\xi^{a}$, where $\sigma\in\Bbb{R}$, to first-order in $\sigma$ we have
\bqa
\dot{x}^{a}&\rightarrow&\dot{\bar{x}}=\dot{x}^{a}+\sigma\xi^{a}_{,c}x^{c}, \\
g_{ab}&\rightarrow&\bar{g}_{ab}=g_{ab}+\sigma g_{ab,c}\xi^{c}, \\
A_{a}&\rightarrow&\bar{A}_{a}=A_{a}+\sigma A_{a,c}\xi^{c}.
\eqa
The action $S[x^{a},e]$ becomes then
\bqa
\bar{S}&=&\int d\lambda[\fr{1}{2}e^{-1}(\dot{x}^{a}+\sigma \xi^{a}_{,c}x^{c})(\dot{x}^{b}+\sigma \xi^{b}_{,c}x^{c})(g_{ab}+\sigma g_{ab,c}\xi^{c}) \nonumber \\
 && -\fr{1}{2}m^{2}e-q(\dot{x}^{a}+\sigma \xi^{a}_{,c}x^{c})(A_{a}+\sigma A_{a,c}\xi^{c})] \nonumber \\
&=&S+\fr{\sigma}{2}\int d\lambda\left[e^{-1}\dot{x}^{a}\dot{x}^{b}({\mathcal L}_{\xi}g)_{ab}-2({\mathcal L}_{\xi}A)_{c}\xi^{c}\right] \nonumber \\
&=& S. \, \blacksquare
\eqa


\begin{thebibliography}{99}

\bibitem{tolman34}
R. C. Tolman, Proc. Nat. Acad. Sci. USA {\bf 20}, 410 (1934).

\bibitem{bondi48}
H. Bondi, Mon. Not. Astron. Soc. {\bf 107}, 343 (1948).

\bibitem{oppenheimer&snyder39}
J. R. Oppenheimer and H. Snyder, Phys. Rev. {\bf 56}, 455 (1939).

\bibitem{newtonian}
This also happens in the Newtonian collapse of homogeneous spherical shells, and it is due to the homogeneity of the matter distribution and its spherical symmetry: homogeneity implies that $m(r)\propto r^{3}$ and spherical symmetry guarantees that the gravitational force per unit mass at radius $r$ is $F\propto mr^{-2}$. Since the equation of motion for a unit test mass is $\ddot{r}=F\propto r$, it follows---from dimensional analysis alone---that the collapse time will be {\em independent} of the radial coordinate $r$. Outer shells are more accelerated (because the mass inside them is larger) than inner ones, and thus collapse faster to travel a larger distance in {\em exactly} the same time needed for inner, less accelerated shells.

\bibitem{yodzis&seifert&hagen73}
P. Yodzis, H. J. Seifert, and H. M. zum Hagen, Comm. Math. Phys. {\bf 34}, 135 (1973).

\bibitem{newman86}
R. P. A. C. Newman, Class. Quantum Grav. {\bf 3}, 527 (1986).

\bibitem{nolan00}
B. C. Nolan, Phys. Rev. D {\bf 60}, 02014 (1999).

\bibitem{papapetrou&hamoui67}
A. Papapetrou and Hamoui, Ann. Inst. Henr\'{\i} Poincar\' e {\bf VI}, 343 (1967).

\bibitem{novikov67}
I. D. Novikov, Sov. Astron. {\bf 10}, 731 (1967).

\bibitem{vickers73}
P. A. Vickers, Ann.Inst. Henr\'{\i} Poincar\' e {\bf 43}, 137 (1973).

\bibitem{ori90}
A. Ori, Class. Quantum Grav. {\bf 7}, 985 (1990).

\bibitem{ori91}
A. Ori, Phys. Rev. D {\bf 44}, 2278 (1991).

\bibitem{riess98}
A. G. Riess {\em et al.}, Astron. J. {\bf 116}, 1009 (1998).

\bibitem{perlmutter98}
S. Perlmutter {\em et al.}, Astrophys. J. {\bf 517}, 565 (1999).

\bibitem{zehani&dekel99}
I. Zehani and A. Dekel, Nature {\bf 401}, 252 (1999).

\bibitem{mellor&moss90}
F. Mellor and I. G. Moss, Phys. Rev. D {\bf 41}, 403 (1990).

\bibitem{lopez95}
C. L\' opez, Gen. Rel. Grav. {\bf 27}, 85 (1995).

\bibitem{humphreys&maartens&matravers98}
N. P. Humphreys, R. Maartens, and D. R. Matravers, ``Regular Spherical Dust Spacetimes'', gr-qc/9804023.

\bibitem{joshi&dwivedi93}
P. S. Joshi and I. H. Dwivedi, Phys. Rev. D {\bf 47}, 5357 (1993).

\bibitem{lake92}
K. Lake, Phys. Rev. Lett. {\bf 68}, 3129 (1992).

\bibitem{lemos92}
J. P. S. Lemos, Phys. Rev. Lett. {\bf 68}, 1447 (1992).

\bibitem{singh96}
T. P. Singh, ``Gravitational Collapse and Cosmic Censorship'', gr-qc/9606016.

\bibitem{dwivedi&joshi92}
I. H. Dwivedi and P. S. Joshi, Class. Quantum Grav. {\bf 9}, L69 (1992).

\bibitem{abramovitz&stegun64}
M. Abramowitz and I. A. Stegun, {\em Handbook of Mathematical Functions} (Dover, New York, 1977).

\bibitem{MTW}
C. W. Misner, K. S. Thorne, and J. A. Wheeler, {\em Gravitation} (W. H. Freeman, San Francisco, 1973), Sec. 32.2.

\bibitem{bonnor62}
See, e.g., W. B. Bonnor, in {\em Recent Developments in General Relativity} (Pergamon, New York, 1962).

\bibitem{brink&vecchia&howe76}
The einbein $e(\lambda)$ is a function of the affine parameter that acts as a gauge field in one dimension for local reparametrizations of the worldline. By definition, it transforms under reparametrizations as $e(\lambda)\rightarrow e'(\lambda')=e(\lambda)d\lambda/d\lambda'$ (which is the transformation rule for the square root of a one-dimensional metric---hence the name `einbein'). The introduction of the einbein in the action (\ref{action}) has the advantage of (i) making it reparameterization invariant, and (ii) obtaining the $m\rightarrow0$ limit (massless particle) in a trivial manner. The einbein action formulation contains the lagrangian and hamiltonian action principles in one unified framework. See, e.g., L. Brink, P. di Vecchia, and P. Howe, Phys. Lett. {\bf 65B}, 471 (1976).

\end{thebibliography}
\end{document}